# Phonon Dispersion Relation, High-Pressure Phase Stability and Thermal Expansion in YVO$_4$

R. Mittal[1,2], M. K. Gupta[1], Baltej Singh[1,2], L. Pintschovius[3,4], Yu D. Zavartsev[5] and S. L. Chaplot[1,2]
[1]Solid State Physics Division, Bhabha Atomic Research Centre, Mumbai 400085, India
[2]Homi Bhabha National Institute, Anushaktinagar, Mumbai 400094, India
[3]Institute for Solid State Physics, Karlsruhe Institute of Technology, 76131, Karlsruhe, Germany
[4]Laboratoire Leon Brillouin, CEA-Saclay, F-91191 Gif sur Yvette Cedex, France
[5]Prokhorov General Physics Institute of the Russian Academy of Sciences, Moscow 119991, Russia

The orthovanadates are useful as host matrices for immobilization of radioactive wastes. The thermodynamic stability of these materials is crucial for their applications in high pressure and temperatures environment. It is necessary to investigate the phonons in the entire Brillouin zone, beyond the zone-centre phonons accessible in previous Raman and infrared experiments. We have carried out extensive neutron inelastic scattering experiments to derive the phonon dispersion relation of YVO$_4$ up to high energy transfer of 65 meV using a single crystal, which are perhaps reported for the first time in any orthovanadate compound. The measured phonon dispersion relation is in good agreement with our first principles density functional theory as well as shell model calculations. The calculated pressure dependence of phonon modes in the zircon and scheelite phases shows unstable modes and violation of the Born stability criteria at high pressure, which may be lead to instability in YVO$_4$ at high pressures. We also calculate large anisotropy in the thermal expansion behavior which arises from difference in anisotropic elasticity and mode Grüneisen parameters.

# I. INTRODUCTION

The zircon structured minerals and vandates are important to a variety of geophysical, geochemical, energy and optical applications[1-22]. The compounds are found in large number of rock types in the Earth's upper mantle. Orthovanadates formed by lanthanides are of interest because of high concentration of various lanthanides present in the waste resulting from reprocessing of light water reactor spent fuel. The compounds are useful as host matrices for immobilization of radioactive wastes. The compounds find wide potential for their application in optoelectronics due to their optical and luminescent properties[3, 15, 23]. At ambient conditions these compounds crystallize[24] in the zircon structure (space group: $I4_1/amd$). The thermodynamic stability of these materials is crucial for their applications at high pressure and temperatures.

The structural stability and thermodynamic properties of $MSiO_4$ (M=Zr, Hf, Th, U) silicates, $RVO_4$ and $RPO_4$ (R= Rare earth) compounds have been systematically investigated [18, 19, 23-30]. Several Raman and infrared studies[1, 5, 18, 31] of the zircon phase at room pressure as well as ab-inito calculations of zone centre modes have been reported. Recently extensive x-ray diffraction, optical and Raman scattering measurements have been carried out under extreme conditions to understand the structural evolution of the ambient pressure zircon phase[2, 18, 23, 31]. The high pressure measurements provided the evidence of the importance of the size of ionic radius of R atom in leading to the sequence of structural phase transitions. The compounds with smaller R ionic radii (like $LuVO_4$) have been observed to undergo an irreversible zircon-to- scheelite (space group: $I4_1/a$, Z = 4) phase transition [28]. However, $CeVO_4$, a compound with a large R atom, revealed zircon to monazite transition[29]. Theoretical ab-inito calculations have been performed to understand the stability of various phases.

High pressure x-ray diffraction studies on $YVO_4$ indicated a zircon to scheelite type phase transition at 8.5 GPa[24]. The scheelite phase is found to be about 10% denser as compared to the zircon phase. On further increase of pressure to about 24 GPa, the scheelite phase undergoes another pressure driven phase transition[32]. A possible structure candidate is the monoclinic fergusonite structure. Raman measurements also indicated that ambient pressure zircon phase transforms to the scheelite phase at 7.5 GPa, while around 24 GPa a reversible second order phase transition occurs[32].

The tetragonal zircon structure has $V^{5+}$ ions tetrahedrally coordinated by oxygen. The space between the isolated $VO_4$ tetrahedral units is occupied by $Y^{3+}$ ions. The Y ions are eight-fold coordinated by oxygen atoms forming $YO_8$ dodecahedral cages. The structure unit can be considered as a chain of alternating edge-sharing $VO_4$ tetrahedra and $YO_8$ dodecahedra extending parallel to the c-axis, with the chain joined along the a-axis by edge- sharing $YO_8$ dodecahedra. For the scheelite phase, $VO_4$ tetrahedra are aligned along the a-axis, whereas along the c-axis $YO_8$ dodecahedra are interspersed between the $VO_4$ tetrahedra.

Phonon properties are important for understanding the thermodynamic behavior of materials under high temperature and pressure conditions. It is necessary to investigate the phonons in the entire Brillouin zone, beyond the zone-centre phonons accessible to Raman and infrared experiments. Earlier, we have shown that transferable interatomic potential models developed for $MSiO_4$ (M=Zr, Hf, Th, U) and $RPO_4$ (R= Rare earth atoms) allow us to calculate [28, 33, 34] the phonon spectra, high pressure and temperature phase diagram as well as thermodynamic properties of these compounds with good accuracy. In particular, we could reproduce the structural phase transformation from the zircon phase to the scheelite phase (body centered tetragonal, *$I4_1/a$*) occurring at high pressure and temperature[24, 28, 35]. Here we present the results obtained from our study on $YVO_4$. We have measured the phonon dispersion relation for $YVO_4$ using the 1T1 neutron spectrometer at LLB, Saclay. We could measure all the phonon branches along the (100) and (001) directions up to 65 meV. The shell model was successfully used for prediction of one-phonon structure factors in order to select most appropriate Bragg points for the measurement of particular phonon branches. The experimental dispersion relation is found to be in satisfactory agreement with the predictions of the model. The extensive measurements as being reported for $YVO_4$ are often not available in similar compounds despite their importance. Further, we have carried out density-functional theory based ab-initio calculation of the phonon spectrum in both the zircon and the scheelite phases. The pressure dependence of the zone-centre modes has been used to identify soft phonon modes and violation of Born stability criteria that may be associated with various phase transitions in $YVO_4$. Further, the calculations successfully reproduced the observed large anisotropic thermal expansion behaviour. In addition we also report phonon calculation in the proposed monoclinic phase (space group I2/a (No 15)) at high pressure, which are useful to provide assignment of reported experimental data.

Section II describes the experimental technique. The lattice dynamic calculations are outlined in section III, while the results and discussion and the conclusions are presented in sections IV and V, respectively.

## II. EXPERIMENTAL

The phonon dispersion relation for YVO$_4$ was measured using the 1T1 triple axis spectrometer at the Laboratoire Léon Brillouin, Saclay. This instrument is equipped with vertically and horizontally focusing monochromators and analyzers resulting in high neutron intensity. A single crystal of about 4 cc was cooled to 10 K in a helium cryostat. Data were collected in the (100)-(010) and (100)-(001) scattering planes. The measurements in the low energy range up to 25 meV were done with pyrolitic graphite (PG002) as monochromator and analyzer, while for higher energies copper (Cu220) was used as monochromator. We could measure almost all the phonon branches along the (100) and (001) directions up to 65 meV.

## III. LATTICE DYNAMICAL CALCULATIONS

The lattice dynamics calculations have been performed using both the empirical potential as well as *ab-initio* methods. The shell model calculations have been carried using the interatomic potentials consisting of Coulombic and short ranged Born-Mayer type interactions terms, given by

$$V(r) = \frac{e^2}{4\pi\varepsilon_o} \frac{Z(k)Z(k')}{r} + a\exp\left[\frac{-br}{R(k)+R(k')}\right] - \frac{C}{r^6} \qquad (1)$$

where, r is the separation between the atoms of a type k and k', and R(k) and Z(k) are, respectively, the effective radius and charge of the k$^{th}$ atom. As in earlier studies, a= 1822 eV and b=12.364 have been treated as constants. This choice has been successfully used earlier to study the phonon density of states and thermodynamic properties of several complex solids. The van der Waals interaction has been introduced only between the oxygen atoms. The stretching potential between V-O bond is of the form

$$V(r) = -D\exp[-n(r - r_o)^2/(2r)] \qquad (2)$$

The parameters used in our calculations are are Z(Y)=2.971, Z(V)= 3.013, Z(O)= -1.496, R(Y)=2.2 Å, R(V)= 0.953 Å, R(O)= 1.855 Å, C = 21.61 eV Å$^6$, D= 1.96 eV, n= 27.24 Å$^{-1}$, r$_o$ =1.706 Å. The polarizibility of the oxygen atom has been introduced in the framework of the shell model[36] with the shell charge Y(O)=-2.46 and shell-core force constant K(O)=60 eVÅ$^{-2}$. The shell model calculations are performed using the software [37] developed at Trombay.

Density functional theory (DFT) has been well known to describe the structural and phonon properties of materials. Vienna ab initio simulation package (VASP-5.4) [38, 39] along with the PHONOPY package has been used for the phonon calculations. The calculations are performed using the projected augmented wave (PAW) formalism of the Kohn-Sham density functional theory within generalized gradient approximation (GGA) for exchange correlation following the parameterization by Perdew, Becke and Ernzerhof[40, 41]. The plane wave pseudo-potential with a energy cutoff of 900 eV was used. The integration over the Brillouin zone has been performed using a k-point grid of 4×4×4, generated using the Monkhorst-Pack method[42]. The criteria for the convergence of total energy and ionic forces were set to $10^{-8}$ eV and $10^{-5}$ eV Å$^{-1}$, respectively. The zircon and scheelite phases of YVO$_4$ both have 24 atoms in a unit cell. A 2×2×2 super cell (192 atoms) of the unit cell has been used for the phonon calculation. We have used the frozen phonon method to calculate the phonon frequencies in entire Brillouin zone. In this approach total energies and inter-atomic forces were calculated at different atomic configurations results from displacement of symmetrically inequivalent atoms along the three Cartesian directions (±x, ±y and ±z). These forces are used to construct the force constants matrix and subsequent phonon frequencies calculation in entire Brillouin zone using PHONOPY software[43].

In order to calculate the pressure dependence of Raman modes in the ziron, scheelite and fergusonite phase of YVO$_4$, we have used the density functional peturbation theory (DFPT) method[44] implemented in VASP. The results obtained from either method are identical at ambient pressure. As the crystal structure of the Fergusonite phase of YVO$_4$ is not known, in our DFT calculation we have started with the known structure of LuVO$_4$[33] and replaced the Lu atom with Y atom, and then relaxed the structure to minimize the total energy.

## IV. RESULTS AND DISCUSSION

### A. Group Theoretical Analysis and Zone Centre Modes in Zircon Phase

The group theoretical analysis of phonon dispersion relation in the zircon phase have been carried out using standard techniques. The group theoretical information have been exploited in diagonalizing the dynamical matrix for wave vectors along the three high symmetry directions. The analysis is useful to distinguish between the various branches and to interpret the experimental results discussed below. The symmetry decomposition of phonon branches in the zircon phase at Γ point, and along the Δ and Λ directions is as follows:

Γ : $2A_{1g} + A_{2g} + A_{1u} + 4A_{2u} + 4B_{1g} + B_{2g} + B_{1u} + 2B_{2u} + 5E_g + 5E_u$ ($E_g$ and $E_u$ modes are doubly degenerate)

Δ : $11\Delta_1 + 7\Delta_2 + 11\Delta_3 + 7\Delta_4$

Λ : $6\Lambda_1 + 2\Lambda_2 + 6\Lambda_3 + 2\Lambda_4 + 10\Lambda_5$ ($\Lambda_5$ modes are doubly degenerate)

The zone center, $A_{1g}$, $B_{1g}$, $B_{2g}$ and $E_g$ modes are Raman active. The $A_{2u}$ and $E_u$ modes are polar exhibiting LO-TO splitting, with their macroscopic field, respectively parallel and perpendicular to the tetragonal *c*-axis. The $B_{1u}$, $A_{2g}$, $B_{1u}$ and $B_{2u}$ modes are optically inactive, but can be observed using inelastic neutron scattering.

The interatomic potential model for $YVO_4$ compounds has been developed based on our experience with $MSiO_4$ and $RPO_4$ compounds[19, 28, 33, 34]. The most detailed check of such a model can be done by calculating the phonon dispersion relation and to compare it with experimental data. A comparison of the calculated long wavelength phonon modes along with the results from our neutron experiments and the reported [23, 32, 45] Raman and infrared data is given in Table II. The data from the neutron experiments up to about 65 meV match quite well with those from the optical experiments. It can be seen that zone centre V-O stretching modes around 100 meV are better described by ab-initio calculations in comparison to the shell model. Further, one of the Eu(LO) mode observed experimentally at around 37.9 meV is significantly overestimated at 48.1 meV from the shell model calculations. The shell model and ab-initio calculated long-wavelength results show a overall satisfactorily agreement with the experimental data with an average deviation of 7% and 4.5% respectively. The calculated ab-initio results are found to matches very well with the experimental data. Further calculations of phonon and thermodynamic behavior in the zircon phase as well as scheelite phase, and phonon calculations in the ferguaonite phase are performed using ab-initio method.

**B. Measurement of Phonon Dispersion Relation**

The scattering vector (**Q**) in inelastic neutron scattering from one phonon excitation is expressed as **Q**=**G**±**q**, where **G** is a Bragg point and **q** is phonon wave vector. We planned the measurements of phonon dispersion relation for **q** along [100] and [001] directions. A lattice-dynamical shell model as described above is first used for the calculation of one-phonon structure factors for the measurements in the *a-b* and *a-c* scattering planes for all the Bragg points in the **Q** range of 2-8 Å$^{-1}$. These calculations show that, in the a-c plane the phonons can be measured in the group theoretical representations $\Delta_1$ and $\Delta_3$ along [100] direction, and in the $\Lambda_1$, $\Lambda_3$ and $\Lambda_5$ along [001] direction. Further in the a-b plane the phonons in the group theoretical representations $\Delta_1$ and $\Delta_4$ along [100] direction can be measured around the Bragg point ($h,k$,0) ($h, k$=even), while the data in the $\Delta_2$ and $\Delta_3$ representations along [100] direction can be measured with ($h,k$,0) ($h, k$=odd). These selection rules arise due to the fact that either the atomic vibrations are not in the scattering plane or phases due to various atoms cancel with each other. These calculations are useful for assignment of observed peaks in the neutron inelastic experiments to specific phonons branches.

We have used (006), (008), (400), (103) and (301) Bragg points for the measurements in representations $\Delta_1$ and $\Delta_3$ along [100], and (003),(005) (0 0 6), (004), (008), (4 0 0), (300), (301), (600) (8 0 0), (802), (702), (207), (208), (205), (500), (1 0 6), (6 0 4), (103), (501), (5 0 3) (601) (701), (403) and (4 0 1) Bragg points for measurements in representations $\Lambda_1$, $\Lambda_3$, and $\Lambda_5$ along [001] directions. Some phonons in representations $\Delta_1$ and $\Delta_4$ along [100] direction are measured using Bragg points (4 0 0 ), (4 2 0), (4 4 0), (6 0 0), (6 2 0), (6 4 0) and (8 2 0) while the Bragg points (5 1 0), (5 3 0), (5 5 0), (7 3 0) and (7 1 0) are used to obtain phonons in $\Delta_2$ and $\Delta_3$ representations. The choice of these Bragg points were made from the prediction of one-phonon structure factors as obtained from shell model calculations. The typical neutron inelastic scattering scans performed in the experiments are shown in Fig. 2. It can be seen that we could get clear phonon peaks in various scans. The experimental scans are fitted to Gaussian functions. The peak positions as obtained from the fitting are assigned to phonon dispersion relation in various group theoretical representations along [100] and [001] directions. The intensities of the phonon peaks as found in the experiments are found to be in qualitative agreement with the calculated one-phonon structure factors. This may be considered encouraging due to the many corrections involved in the experimental intensities and the predictions based on an empirical shell model. The large size of the crystal as well as the high neutron flux at the sample position have enabled to measure almost all the phonon dispersion relation upto 65 meV (Fig. 3) along both the directions. The experimental data matches very well with the ab-initio calculations as well as shell model.

## C. Phonon Density of States in Zircon and Scheelite Phases

The ab-initio calculated phonon density of states in zircon and scheelite phases show (Fig. 4) that in both the phases Y atoms mainly contribute in the low energy range up to 40 meV. However, the contribution from the V and O atoms in zircon and scheelite phase is up to 120 and 110 meV respectively. Normally one expects that in the high pressure phase the decrease in volume would lead to the increase in the range of the phonon spectra. However in $YVO_4$, we find that V-O bond length in the scheelite phase increases to 1.743 Å in comparison to the value of 1.728 Å in the zircon phase. This mainly results in the decrease in the range of vibrational spectrum in the scheelite phase. The band gap in the phonon spectrum in the zircon phase is from 62 to 95 meV, while in the scheelite phase the increase in V-O bond length shifts the band gap to 60 - 85 meV. The calculated Born effective charges of various atoms are given in Table III. There is little difference in the ionicity of atoms in both the phases. Therefore, it appears that the difference in the range of phonon spectra of the two phases is mainly due to the changes in the structures.

## D. Calculated Elastic Constants Born Stability Criteria in Zircon and Scheelite Phases

Ultrasonic studies have been performed to experimentally[46] determine the elastic constants in the zircon phase. The calculated values (Table IV) match very well with the experimental data. The small difference in the experiment and calculation elastic constants values is also reflected in the slopes corresponding to various longitudinal and transverse acoustic branches in the neutron data and ab-initio calculations (Fig. 3). The value of longitudinal elastic constants $C_{33}$ in the zircon phase is about 30 % larger in comparison to $C_{11}$, while in the scheelite phase both the longitudinal elastic constants ($C_{11}$ and $C_{33}$) have same value. Further we find that there is large difference in the values of $C_{33}$, $C_{66}$ and $C_{12}$ in the zircon and scheelite phases. The calculated bulk modulus (Table IV) in both the phases matches very well with the experimented values. As expected the calculated bulk modulus in the high pressure scheelite phase is larger (~ 11 %) in comparison to that in the zircon phase.

The zircon and scheelite phases of $YVO_4$ are known to exhibit phase transitions on increase of pressure at 8 GPa and 24 GPa respectively. We have calculated pressure dependence of Born stability conditions in both the phases upto 10 GPa and 30 GPa respectively. Under hydrostatic pressure, for a stable tetragonal structure the following Born stability conditions should hold

$$C_{44}-P>0,\ C_{66}-P>0,\ C_{11}-C_{12}-2P>0,\ (C_{33}-P)(C_{11}+C_{12})-2(C_{13}+P)^2>0$$

The calculated pressure dependence of Born stability criteria in both the phases is shown in Fig. 5. The stability conditions related to $C_{44}$ and $C_{66}$ show large pressure dependence. It can be seen that in the zircon and scheelite phase, the Born stability criteria related to $C_{66}$ elastic constant is found to be violated at about 10 GPa and 30 GPa respectively, indicating instability of both the structures. The $C_{66}$ elastic coinstant is related to the transverse acoustic branch in the (110) direction with vibrations in the a-b plane.

**E. High Pressure Behaviour**

The zircon phase is known to undergo phase transition to scheelite phase at about 8 GPa. We have calculated the enthalphy, $H(T,P) = \phi(T,V) + PV$, where $\phi$ is the internal energy, and *P, V* and *T* are pressure, volume and temperature respectively, in both the zircon and scheelite phases of YVO$_4$. The calculated difference in enthalpy (ΔH) between the zircon and scheelite phases shows (Fig. 6 (b)) that scheelite is stable above 5 GPa. The calculated pressure dependence of unit cell volume (Fig. 6 (a)) shows a drop of about 11 % at the zircon to scheelite phase transition in agreement with the experimental data. These calculations indicate that the transition to scheelite phase is a first order transition and is in agreement with the previous calculations[32].

Further, we have calculated pressure dependence of Raman active modes (Fig. 7) in zircon phase. The calculated Raman modes are assigned to various Group theoretical representations. All the Raman active modes show normal behavior and their energies increase with increase in pressure. However we find that the energy of optically inactive $B_{1u}$ mode continues to decrease with increase of pressure (Fig. 8). According to Landau theory of phase transitions the energy E of a soft phonon mode follows the relation $E^2 = A (P - P_0)$ (where A is constant and $P_0$ is the pressure at which mode become unstable) near the phonon instability. As shown in Fig. 8, $E^2$ decreases linearly with increase of pressure, with $P_0$ =8.5 GPa. Since zircon to scheelite transition at 8 GPa is of first order in nature with decrease in volume of about 11 %, the phase transition may not be only due to the softening of the mode. However, the unstable $B_{1u}$ mode may trigger the instability in the lattice. As shown in Fig. 9, we have plotted the displacement pattern of the unstable $B_{1u}$ mode, which indicates rotation of VO$_4$ tetrahedra about the c-axis.

The scheelite phase is known to undergo second order phase transition at about 24 GPa. The calculated volumes and the enthalpy difference between the scheelite and fergusonite phases indicate that there is no discontunity in the P-V or ΔH vs P diagrams. This confirms the results of the previous calculations[32] that transition of scheelite to fergusonite is of second order in nature. High pressure Raman measurements have been reported in the scheelite phase. The symmetry decomposition of phonon modes in the scheelite phase is as follows:

Γ : $3A_g + 5A_u + 5B_g + 3B_u + 5E_g + 5E_u$ ($E_g$ and $E_u$ modes are doubly degenerate)

Except for the phonons of $B_u$ symmetry, all others are either Raman ($A_g$, $B_g$ and $E_g$) or infrared ($A_u$ and $E_u$) active. The comparison of the calculated and experimental pressure dependence of Raman active phonon modes in scheelite phase is shown in Fig. 10. We find that the energy of one of the $B_g$ mode decreases with increase in pressure. Further we have also calculated (Fig. 11) the pressure dependence of modes in the fergusonite phase. Ref. [32] reports the assignments of Raman modes in the zircon and scheelite phases. However, no such assignments have been reported in the fergusonite phase. Fig. 12 shows the experimental data for the pressure dependence of $B_g$ mode in scheelite phase up to 24 GPa, and of the lowest energy unassigned mode in the fergusonite phase up to 30 GPa.

Group theoretical analysis in the fergusonite phase indicates that 36 modes at the zone centre can be classified as

Γ: $8A_g + 8A_u + 10B_g + 10B_u$.

The $A_g$ and $B_g$ modes are Raman active while the $A_u$ and $B_u$ modes are infrared active. So the unassigned mode in the Raman experiment (in Fig. 12) may belong to the $A_g$ or $B_g$ symmetry. The calculated pressure dependence of the lowest energy Raman modes of the $A_g$ and $B_g$ symmetry is also shown in Fig. 12. We find that in our calculations the energy of the $B_g$ mode increases with increase in pressure and it qualitatively matches with the experimental data. The calculated slope of the energy of $A_g$ mode with pressure is negative. So we assign the mode in the Raman experiment to the $B_g$ symmetry.

## F. Anisotropic Thermal Expansion

Several efforts have been dedicated to measure the thermal-expansion behavior of YVO$_4$ in the zircon phase. Interferometrry technique has been used to measure the accurate thermal expansion coefficients along the a and c axes. The thermal expansion coefficients have been found to be highly anisotropic. We have calculated the thermal expansion behavior in both the zircon and scheelite phases.

The anisotropic linear thermal expansion coefficients have been calculated within the quasiharmonic approximation as given by:

$$\alpha_l(T) = \frac{1}{V_0} \sum_{q,i} C_V(q,i,T) [s_{l1}\Gamma_a + s_{l2}\Gamma_b + s_{l3}\Gamma_c] , \qquad l = a, b, c$$

Where s$_{ij}$ are elements of elastic compliances matrix (which is the inverse of the elastic constant matrix) at constant temperature (0 K), V$_0$ is volume at 0 K and C$_V$ (q, $i$, T) is the specific heat at constant volume due to $i^{th}$ phonon mode with wave vector **q** in the Brillouin zone. The linear Grüneisen parameters ($\Gamma_l(E_{q,i})$) are defined using the following relation

$$\Gamma_l(E_{q,i}) = \left(-\frac{\partial \ln E_{q,i}}{\partial \ln l}\right)_{T,l'} ; \; l, l' = a, b, c \;\&\; l \neq l'$$

For tetragonal system $\Gamma_a = \Gamma_b$. For calculation of $\Gamma_l(E_{q,i})$ we have calculated the phonon spectra in the entire Brillouin zone at ambient pressure and again with anisotropic stress. An anisotropic stress of 5 kbar is applied in the calculations by changing the lattice constant 'a' and keeping the 'c' parameter constant, and vice versa. The calculated anisotropic mode Grüneisen parameters as a function of phonon energy in the zircon and scheelite phases of YVO$_4$ are shown in Fig. 13.

The calculated $\Gamma_a$ and $\Gamma_c$ as a function of phonon energy as averaged over all the phonons in the Brillouin zone (Fig. 13) in both the phases are highly anisotropic. The $\Gamma_l$ ($l=a,c$) values in both the phases are mostly positive. The maximum positive value of $\Gamma_c$ in the zircon phase is for phonons of about 25 meV. The phonons of energy 15-25 meV in the scheelite phase have maximum positive value of $\Gamma_c$ of about +2. The acoustic phonons below 2 meV have slight negative values of $\Gamma_a$ of about -1. The V-O stretching modes around 100 meV in both the phases have $\Gamma_l$ ($l$=a,c) values of about +2.

The calculated elastic compliance matrix elements ($s_{ij}$) (Table V) along with the $\Gamma_l$ ($l$=a, c) values are used for the calculation of linear thermal expansion coefficients $\alpha_l$ ($l$=a, c) values (Fig. 14) in both the phases. It can be seen that the calculated $\alpha_c$ values in the zircon phase are about three times the $\alpha_a$ values, while in the scheelite phase the ratio of $\alpha_c/\alpha_a$ is about 1.5 The calculated thermal expansion behavior in the zircon phase matches (Fig. 14) very well with the available experimental data [47, 48]. Further, $\alpha_c$ values in both the phases are nearly same. However, $\alpha_a$ values in the scheelite phase are about 2 times in comparison that in the zircon phase. The difference in $\alpha_l$ ($l$=a, c) values in both the phases arises due to the difference in the calculated $s_{ij}$ and $\Gamma_l$ ($l$=a, c) values (Fig. 14) in both the phases. The calculated $\Gamma_l$ ($l$=a,c) values in both the phases for phonon below 2 meV have negative values of up to -1. However, the calculated $\alpha_a$ values have very low negative values (-0.03 $\times 10^{-6}$ K$^{-1}$) due to the insignificant weight of the phonon spectra below 2 meV.

The computed contribution of phonons of energy E, averaged over the entire Brillouin zone (Fig 15), to the thermal expansion coefficient at 300 K shows that phonon modes in the range of 25 to 35 meV contribute maximum to $\alpha_l$ ($l$=a,c) in the zircon phase, while in scheelite phase the contribution is maximum from 15 to 25 meV phonons. The calculated density of states (Fig. 4) shows that the phonons corresponding to the V and O atoms in the same energy range differ significantly in the two phases. We have shown displacement patterns of representative zone centre modes in the two phases around respective energies in Fig. 16. The eigenvector of the $E_u$(TO) at 29.9 meV in the zircon phase shows that Y and V atoms move perpendicular to each other. The contribution to the thermal expansion coefficient due to this mode (assuming it as an Einstein mode with one degree of freedom), along various axes is: $\alpha_a$ = 1.45 $\times 10^{-6}$ K$^{-1}$, $\alpha_c$ = 4.30 $\times 10^{-6}$ K$^{-1}$. On the other hand, in the scheelite phase, the $B_g$ mode at 22.8 meV of has displacement of both the Y and V atoms show along b-axis and contribute $\alpha_a$ = 2.06 $\times 10^{-6}$ K$^{-1}$, $\alpha_c$ = 3.03$\times 10^{-6}$ K$^{-1}$.

## V. CONCLUSIONS

The inelastic neutron scattering measurements on a single crystal of YVO$_4$ have been used to obtain the phonon dispersion relation in the zircon phase. Such extensive measurements on a orthovanadate are perhaps reported for the first time. The ab-initio calculations of the phonon spectrum are performed in the entire Brillouin zone in various phases. The good agreement between the extensive single crystal experiment and the ab-initio calculations is highly satisfactory and thus the ab-initio calculation of the high pressure phases provide a robust model of the orthovanadates. The pressure dependence of phonon modes has been used to reveal soft phonon modes in the zircon and scheelite phases. The soft modes may be associated with the high pressure phase transitions. We have explained the large difference in the thermal expansion behavior of zircon and scheelite phases as well as large anisotropic behavior along a and c-axes. This study may be further useful for a comprehensive study of the thermodynamic stability of other orthovanadates in high-pressure and high-temperature environment likely to be seen in nuclear waste immobilization.

TABLE I. Comparison between the experimental [24, 32] (at 293 K) and calculated structural parameters (at 0 K) of zircon and scheelite phase of $YVO_4$. For zircon structure (body centered tetragonal, $I4_1/amd$) the Y, V and O atoms are located at (0, 0.75, 0.125), (0, 0.25, 0.375) and (0, $u$, $v$) respectively and their symmetry equivalent positions. For scheelite structure (body centered tetragonal, $I4_1/a$) the Y, V and O atoms are located at (0, 0, 0.5), (0, 0, 0) and ($u$, $v$, $w$) respectively and their symmetry equivalent positions.

|   | Experimental (Zircon) | Calculated shell model (Zircon) | Calculated ab-initio (Zircon) | Experimental (Scheelite) | Calculated ab-initio (Scheelite) |
|---|---|---|---|---|---|
| $a$ (Å) | 7.1183 | 7.0434 | 7.2024 | 5.0702 | 5.0685 |
| $c$ (Å) | 6.2893 | 6.5781 | 6.3201 | 11.3253 | 11.3207 |
| $u$ | 0.4342 | 0.4209 | 0.4346 | 0.2550 | 0.2554 |
| $v$ | 0.2008 | 0.2225 | 0.2003 | 0.1440 | 0.1444 |
| $w$ |   |   |   | 0.0804 | 0.0803 |

TABLE II. Comparison of the experimental and calculated zone-centre optic phonon modes in meV units. The experimental data of Raman active modes, infrared active $A_{2u}$ and $E_u$ modes are from Ref [32], Ref [45] and Ref [23] respectively.

|  | Raman and Infra red modes | Present Neutron Experiment | Calculated *Ab-initio* | Calculated (Shell model) |
| --- | --- | --- | --- | --- |
| $A_{1g}$ | 47.0 | 47.2 | 44.5 | 46.2 |
|  | 110.6 |  | 109.3 | 105.8 |
| $A_{2g}$ |  | 22.45 | 21.6 | 20.7 |
| $B_{1g}$ | 19.5 | 19.77 | 18.6 | 21.3 |
|  | 32.2 | 32.7 | 31.4 | 33.6 |
|  | 60.7 | 60.4 | 57.3 | 60.5 |
|  | 101.3 |  | 103.1 | 110.2 |
| $B_{2g}$ | 32.2 | 32.0 | 31.6 | 26.1 |
| $E_g$ |  | 17.56 | 16.3 | 17.3 |
|  | 20.3 | 20.60 | 19.6 | 18.2 |
|  | 32.3 | 32.34 | 29.2 | 33.0 |
|  |  | 46.88 | 45.9 | 50.2 |
|  | 104.1 |  | 103.7 | 104.5 |
| $A_{1u}$ |  | 41.7 | 41.7 | 41.2 |
| $B_{1u}$ |  | 11.44 | 12.5 | 11.0 |
| $B_{2u}$ |  | 55.5 | 53.7 | 53.9 |
|  |  |  | 107.5 | 104.7 |
| $A_{2u}$(LO) | 43.9 |  | 42.6 | 47.2 |
|  |  | 57.4 | 53.3 | 59.9 |
|  | 118.4 |  | 116.2 | 115.0 |
| $A_{2u}$(TO) | 27.9 | 28.6 | 26.8 | 31.0 |
|  | 55.6 | 55.9 | 53.0 | 55.9 |
|  | 100.0 |  | 102.0 | 110.1 |
| $E_u$(LO) | 27.5 | 26.2 | 25.9 | 32.4 |
|  | 38.5 |  | 35.1 | 35.2 |
|  | 39.1 | 37.9 | 36.9 | 48.1 |
|  | 115.4 |  | 112.7 | 107.6 |
| $E_u$(TO) | 23.9 | 24.5 | 22.8 | 24.4 |
|  | 32.5 | 32.3 | 29.9 | 34.4 |
|  | 38.5 | 37.7 | 36.9 | 38.9 |
|  | 96.5 |  | 96.8 | 101.9 |

TABLE III: The Born effective charge tensors of various atoms in unit of *e*.

| Atom | Zircon | | | Scheelite | | | Ferguson ite | | |
|---|---|---|---|---|---|---|---|---|---|
| O1/O2 | -0.76 | 0.00 | 0.00 | -2.68 | -1.07 | -0.88 | -1.29/-3.24 | -0.49/0.90 | -0.62/0.36 |
| | 0.00 | -3.02 | 1.10 | -0.88 | -1.75 | -0.29 | -0.49/0.81 | -1.99/-1.99 | -0.82/-0.5 |
| | 0.00 | 1.26 | -2.31 | -0.94 | -0.22 | -2.00 | -0.52/0.47 | -1.03/-0.27 | -1.24/-1.24 |
| V | 3.43 | 0.00 | 0.00 | 4.05 | 0.63 | -0.01 | 4.32 | 0.02 | 0.84 |
| | 0.00 | 3.43 | 0.00 | -0.63 | 4.05 | 0.05 | 0.00 | 3.79 | 0.00 |
| | 0.00 | 0.00 | 4.21 | 0.000 | 0.00 | 3.71 | 0.13 | -0.04 | 4.54 |
| Y | 4.12 | 0.00 | 0.00 | 4.79 | -0.28 | 0.00 | 4.70 | 0.00 | -0.32 |
| | 0.00 | 4.12 | 0.00 | 0.28 | 4.79 | -0.03 | 0.00 | 4.33 | 0.00 |
| | 0.00 | 0.00 | 5.013 | 0.00 | 0.00 | 4.31 | 0.03 | 0.00 | 4.73 |

TABLE IV. Comparison between the experimental[5, 46] and calculated elastic constants and bulk modulus of YVO$_4$ in zircon phase and calculated elastic constants in scheelite phase. Elastic constants and bulk modulus are given in GPa units. The experimental data of bulk modulus for zircon and scheelite phase of YVO$_4$ are from Refs.[5]. The experimental data are at P=0 and T=300 K, while the calculatins are at P=0 and T=0.

| Elastic constant | Experimental (Zircon) | Calculated ab-initio (Zircon) | Experimental (Scheelite) | Calculated ab-initio (Scheelite) |
|---|---|---|---|---|
| $C_{11}$ | 244.5 | 222.1 | | 223.1 |
| $C_{33}$ | 313.7 | 292.7 | | 198.5 |
| $C_{44}$ | 48.2 | 43.6 | | 52.6 |
| $C_{66}$ | 16.2 | 18.4 | | 63.1 |
| $C_{12}$ | 48.9 | 45.0 | | 114.5 |
| $C_{13}$ | 81.1 | 82.5 | | 95.5 |
| B | 132.3 | 123.6 | 140.4 | 139.7 |

TABLE V. The calculated elements of the elastic compliance matrix in the zircon and scheelite phases of YVO$_4$ at P=0.

| Compliance | Zircon | Scheelite |
|---|---|---|
| $s_{11}$ (TPa$^{-1}$) | 5.1 | 6.95 |
| $s_{33}$ (TPa$^{-1}$) | 4.1 | 6.86 |
| $s_{44}$ (TPa$^{-1}$) | 22.9 | 19.0 |
| $s_{66}$ (TPa$^{-1}$) | 54.4 | 18.4 |
| $s_{12}$ (TPa$^{-1}$) | -0.56 | -3.01 |
| $s_{13}$ (TPa$^{-1}$) | -0.13 | -1.9 |

FIG 1. (Color online) The crystal structure of YVO$_4$ in zircon, scheelite and fergusonite phases. Key- Y: green, O: blue and V: red.

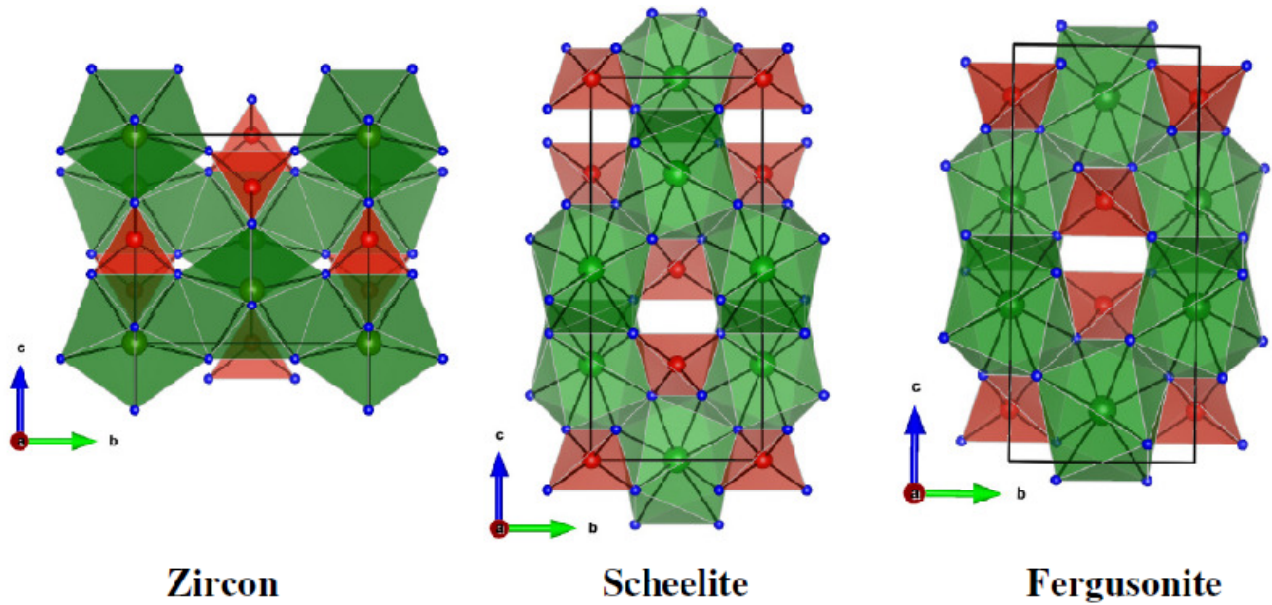

FIG 2. (Color online) Typical neutron inelastic scattering scans performed in the neutron inelastic experiments. Open circles are the experimental data while solid lines are fitted through the experimental data.

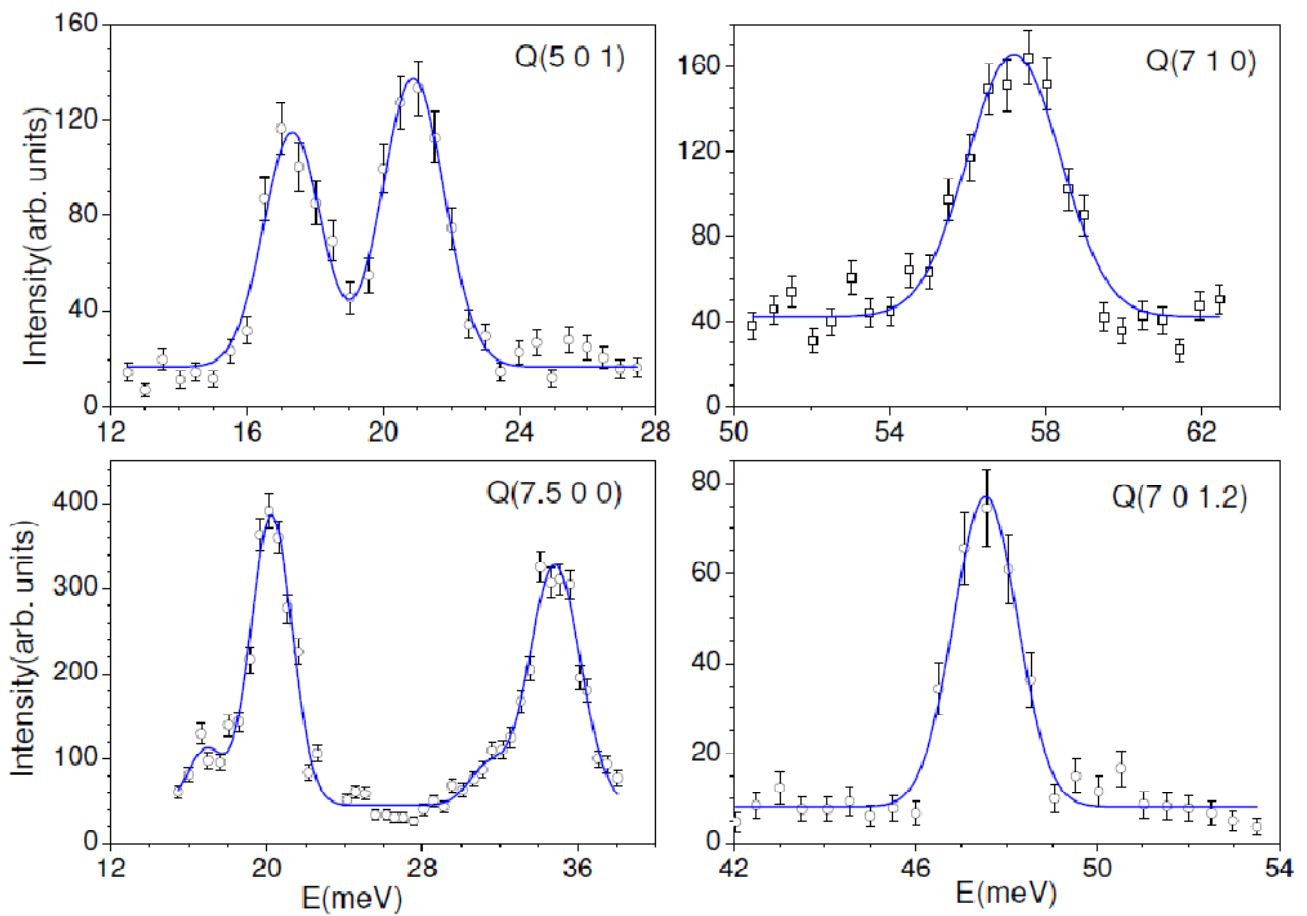

FIG 3. (Color online) The experimental phonon dispersion curves in zircon phase of $YVO_4$ along with the lattice dynamical calculations. The solid circles give the phonon peaks identified in the neutron experiments.

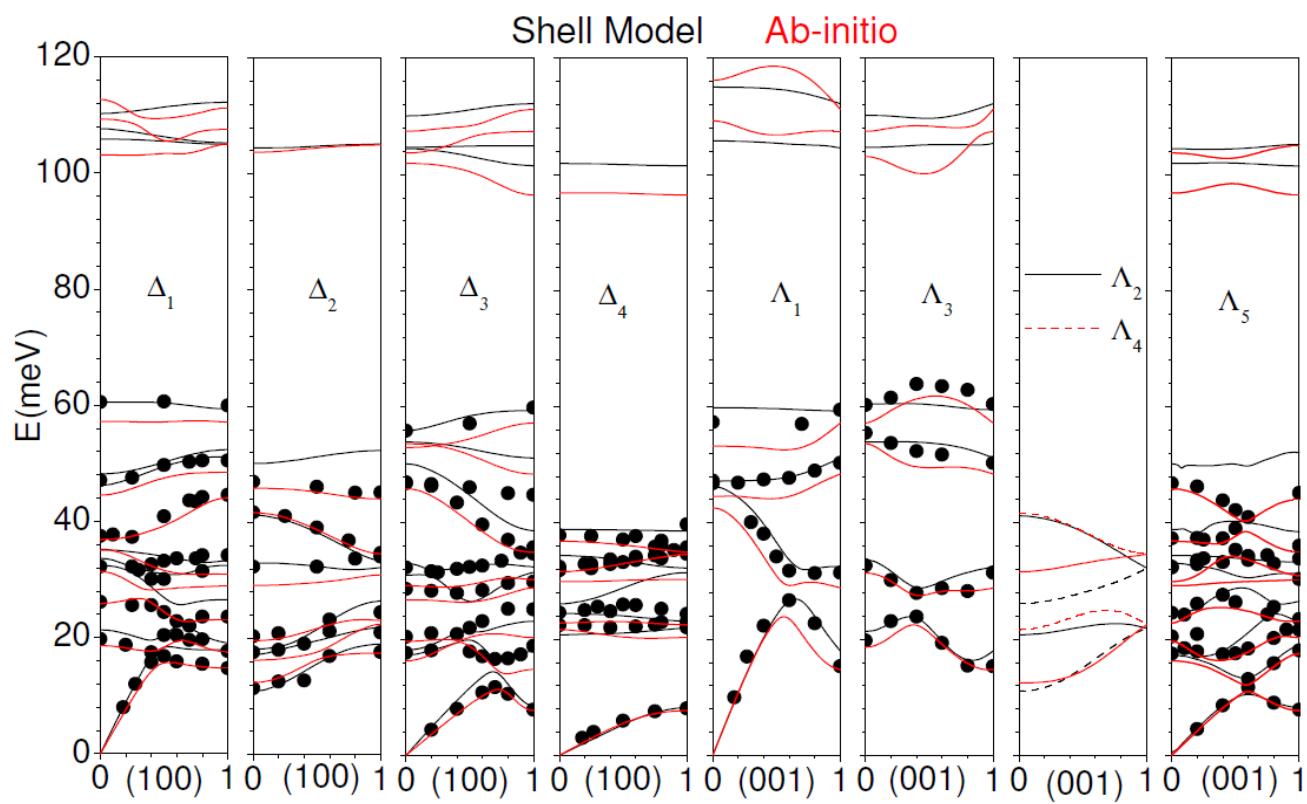

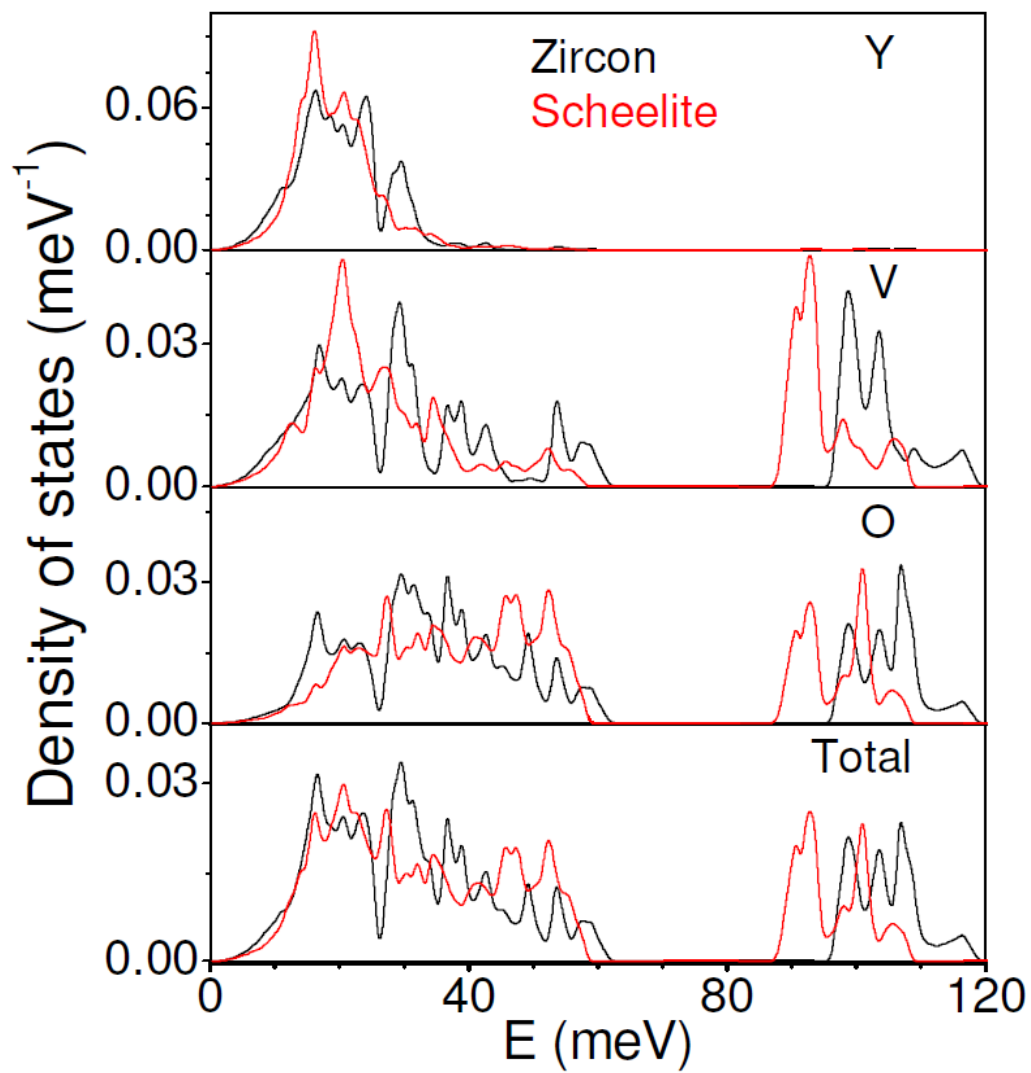

FIG 4. (Color online) The ab-initio calculated partial density of states of various atoms and total density of states in zircon and scheelite phases of YVO$_4$.

FIG 5. The ab-initio calculated pressure dependence of Born stability criteria in zircon and scheelite phase of YVO$_4$.

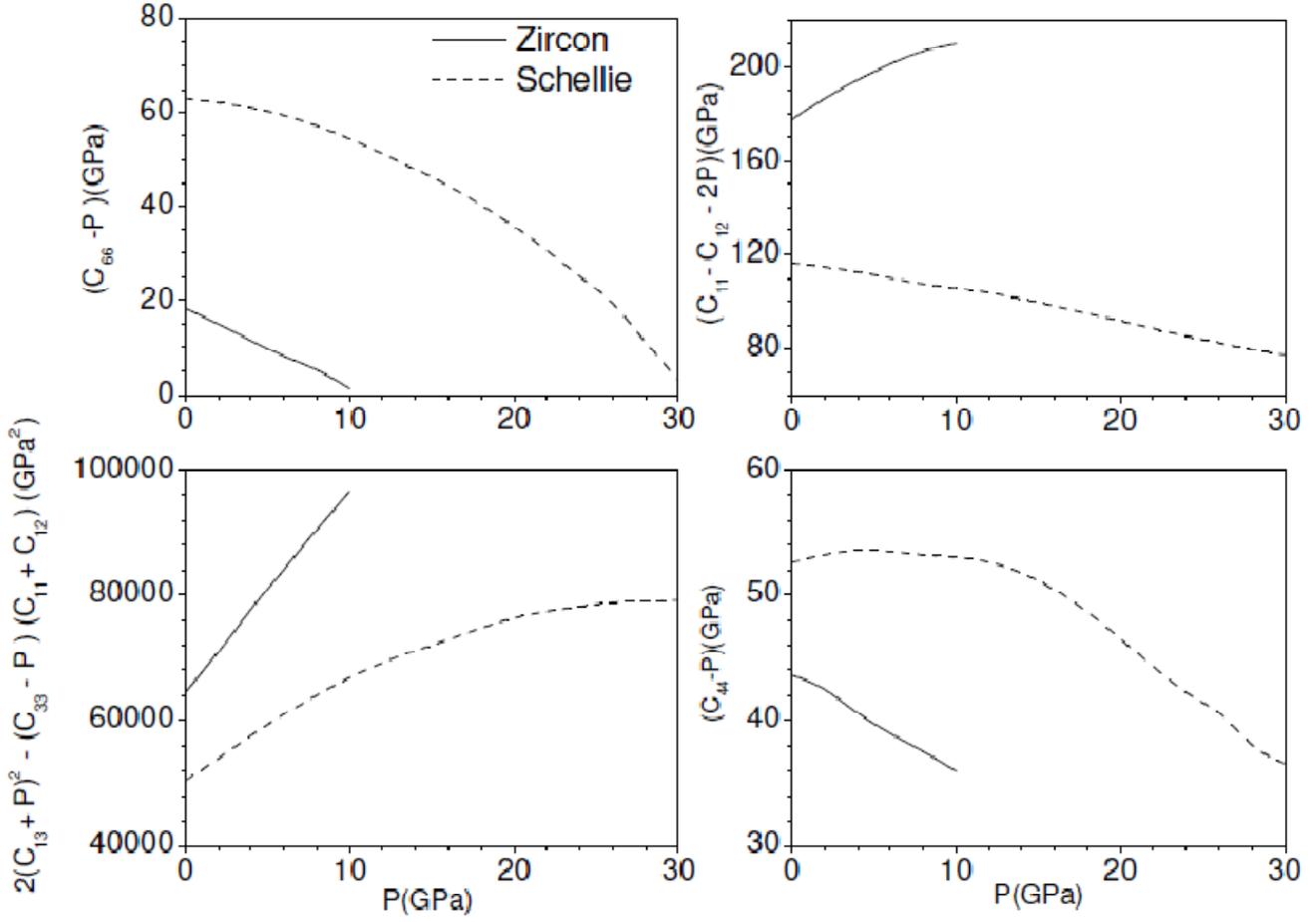

FIG 6. (Color online) (a) The ab-initio calculated unit cell volume of YVO$_4$ in zircon, scheelite and fergusonite phases. (b) The calculated enthalpy difference ($\Delta H$) in the zircon and scheelite phases.

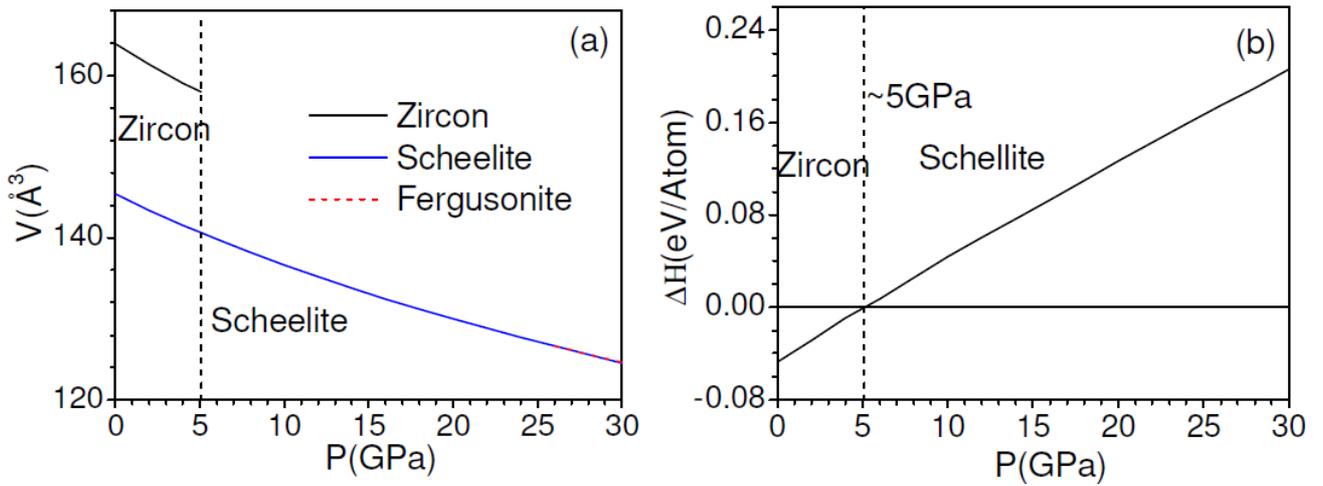

FIG 7. The ab-initio calculated and experimental[32] pressure dependence of Raman modes in zircon phase of YVO$_4$.

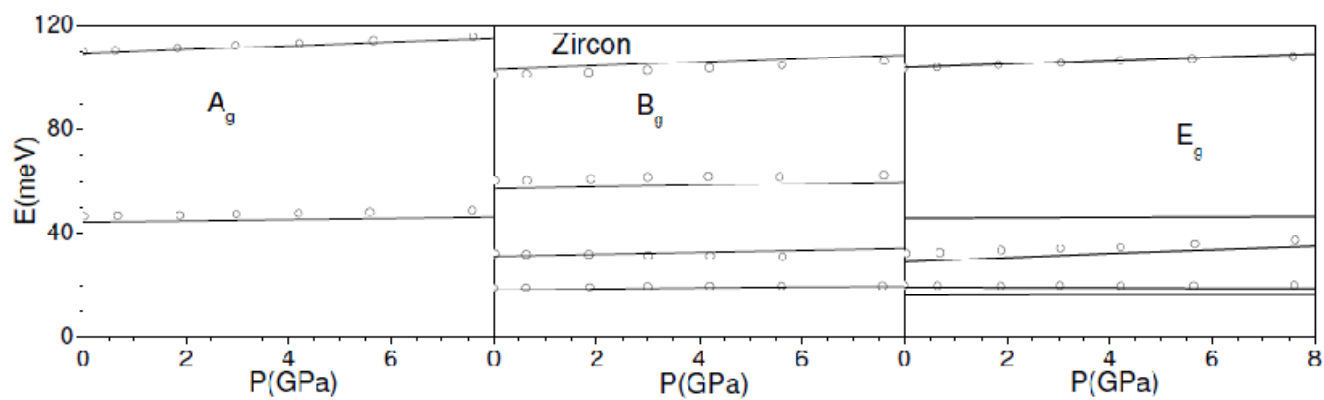

FIG 8. The ab-initio calculated pressure dependence of B$_{1u}$ mode in zircon phase of YVO$_4$.

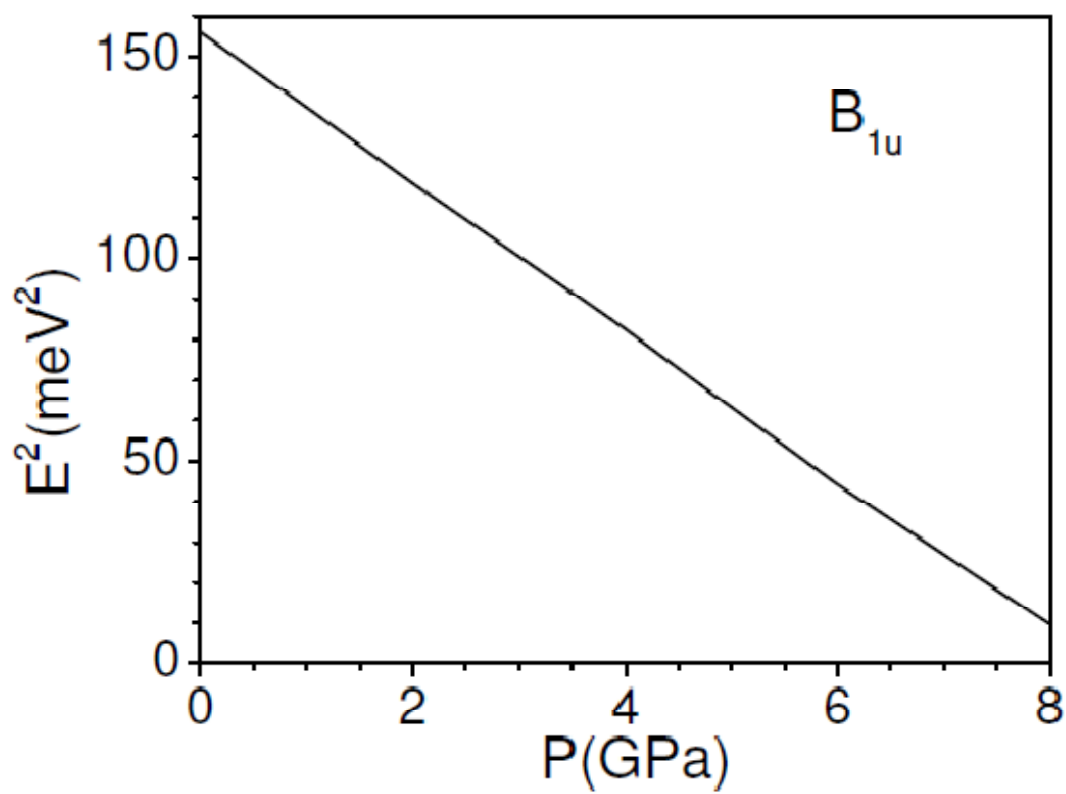

FIG 9 (Color Online) The atomic displacement pattern of $B_{1u}$ Phonon mode (E=12.4 meV) in the zircon phase of $YVO_4$. Key- Y: green, O:blue and V: red.

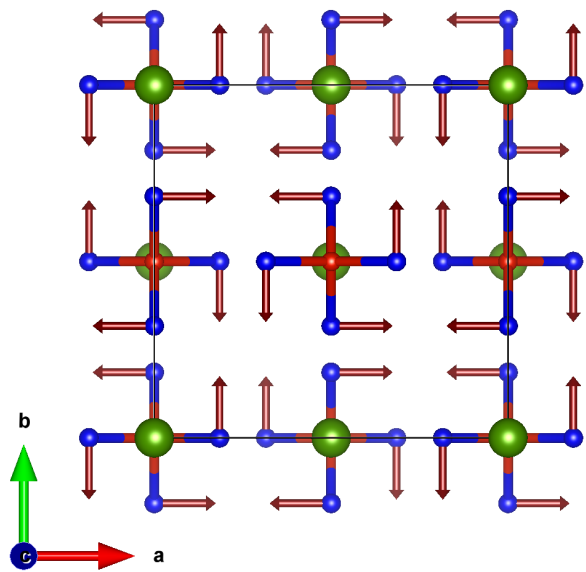

FIG 10. The ab-initio calculated and experimental[32] pressure dependence of Raman modes in scheelite phase of $YVO_4$.

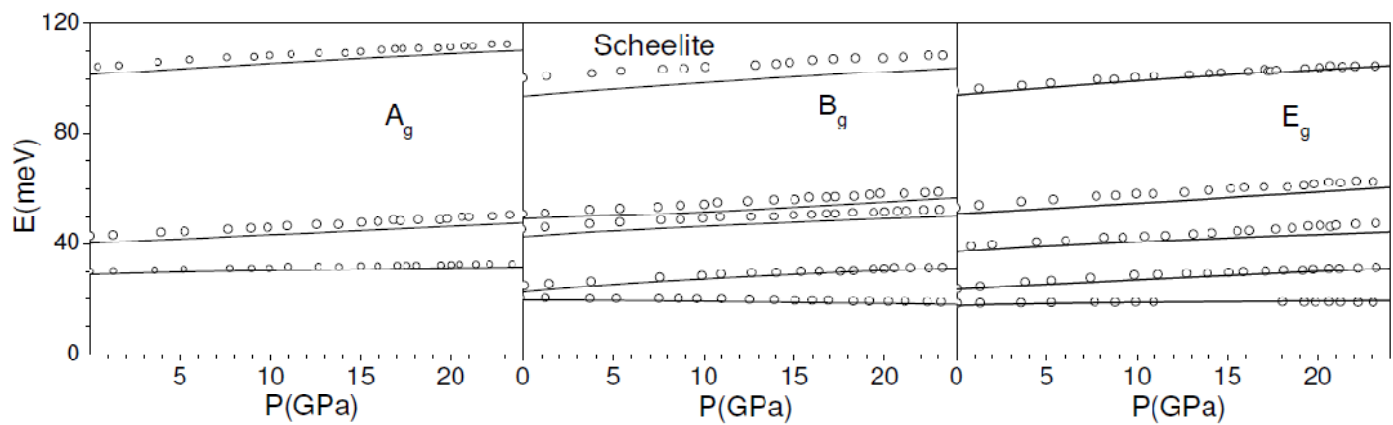

FIG 11. The ab-initio calculated and experimental[32] pressure dependence of Raman modes in ferugsonite phase of YVO$_4$.

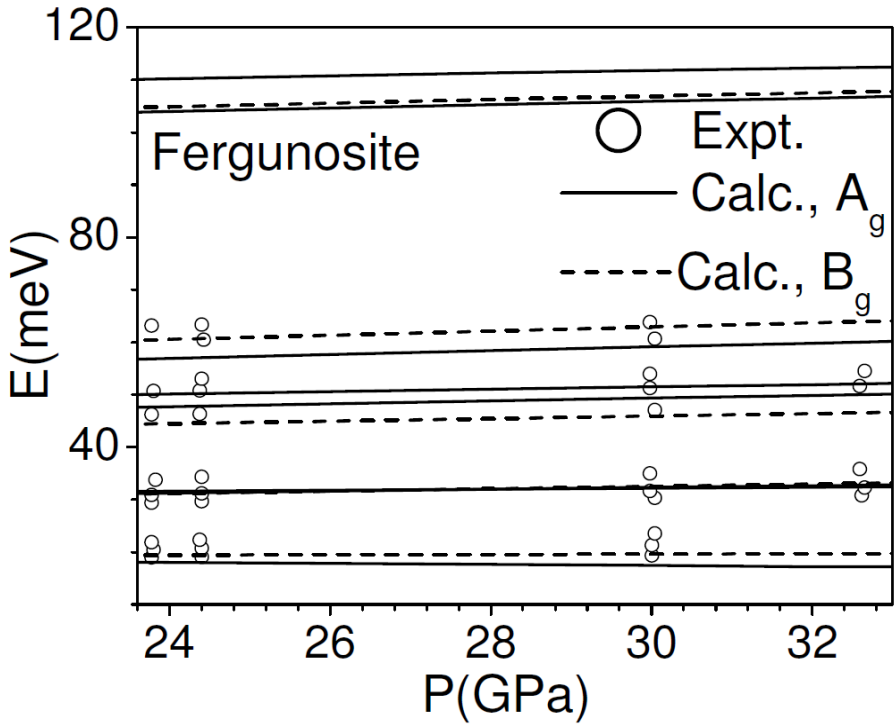

FIG 12. (Color online) The ab-initio calculated and experimental[32] pressure dependence of B$_g$ Raman modes in scheelite. The mode assignment in the experimental data of fergusonite phase is not available. The calculated pressure dependence of two lowest Raman active modes in the fergusonite phase is also shown.

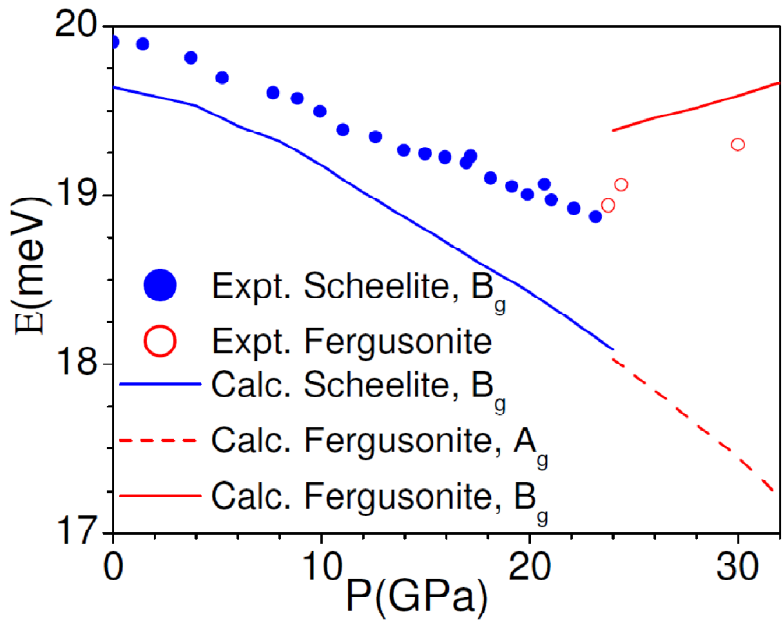

FIG 13. (Color online) The calculated anisotropic mode Grüneisen parameters $\Gamma_a$ and $\Gamma_c$ in zircon and scheelite phase of $YVO_4$.

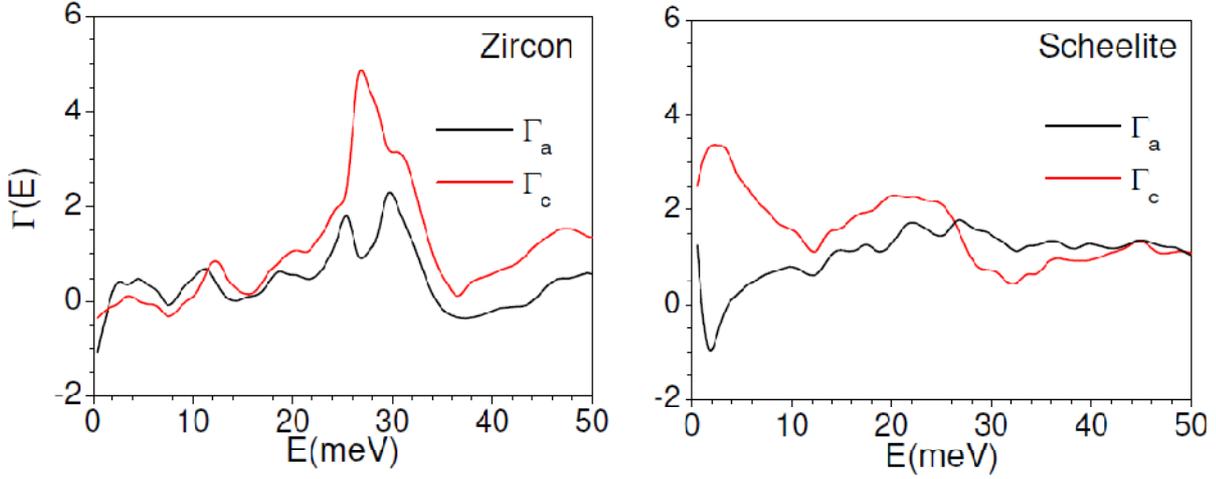

FIG 14. (Color online) The calculated linear thermal expansion behavior ($\alpha_l$ ($l$=a,c)) in zircon and scheelite phase of $YVO_4$. The available experimental data[47, 48] (open and closed circles) from the literature in the zircon phase is also shown.

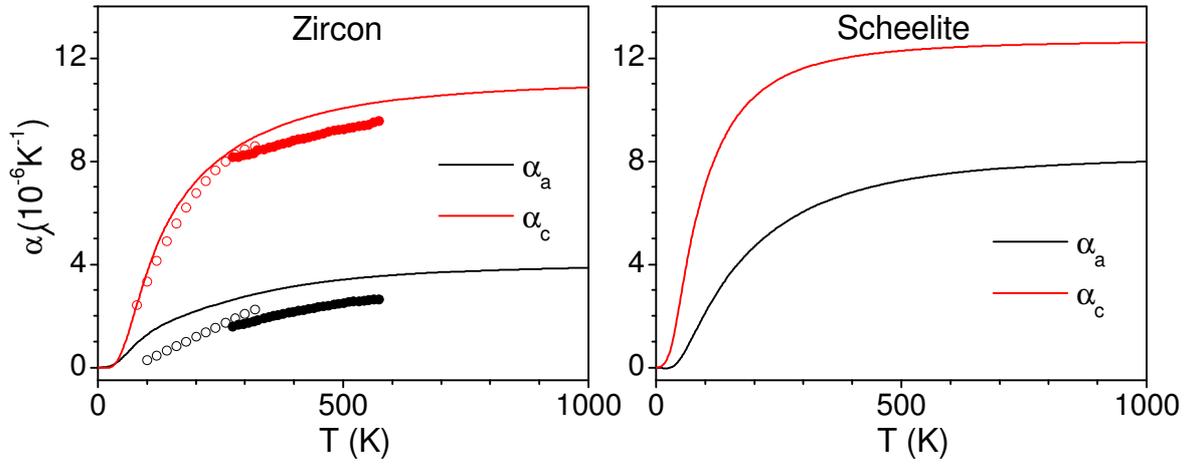

FIG 15. (Color online) The calculated contribution to the volume thermal expansion coefficient from phonons of energy E averaged over the entire Brilloun zone as a function of phonon energy (E) at 300K in the zircon and scheelite phase of YVO$_4$.

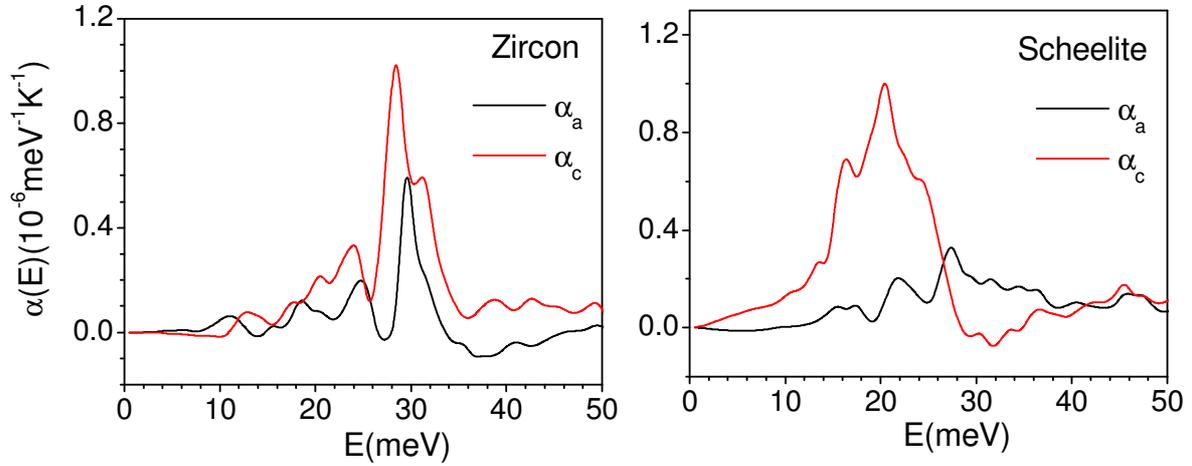

FIG 16. (Color Online) The atomic displacement pattern of selected Phonon modes in the zircon and schellite phase of YVO$_4$. The number below the figure gives the phonon energies, $\Gamma_a$, $\Gamma_c$ and $\alpha_a$ and $\alpha_c$ respectively. Key- Y: green, O:blue and V: red.

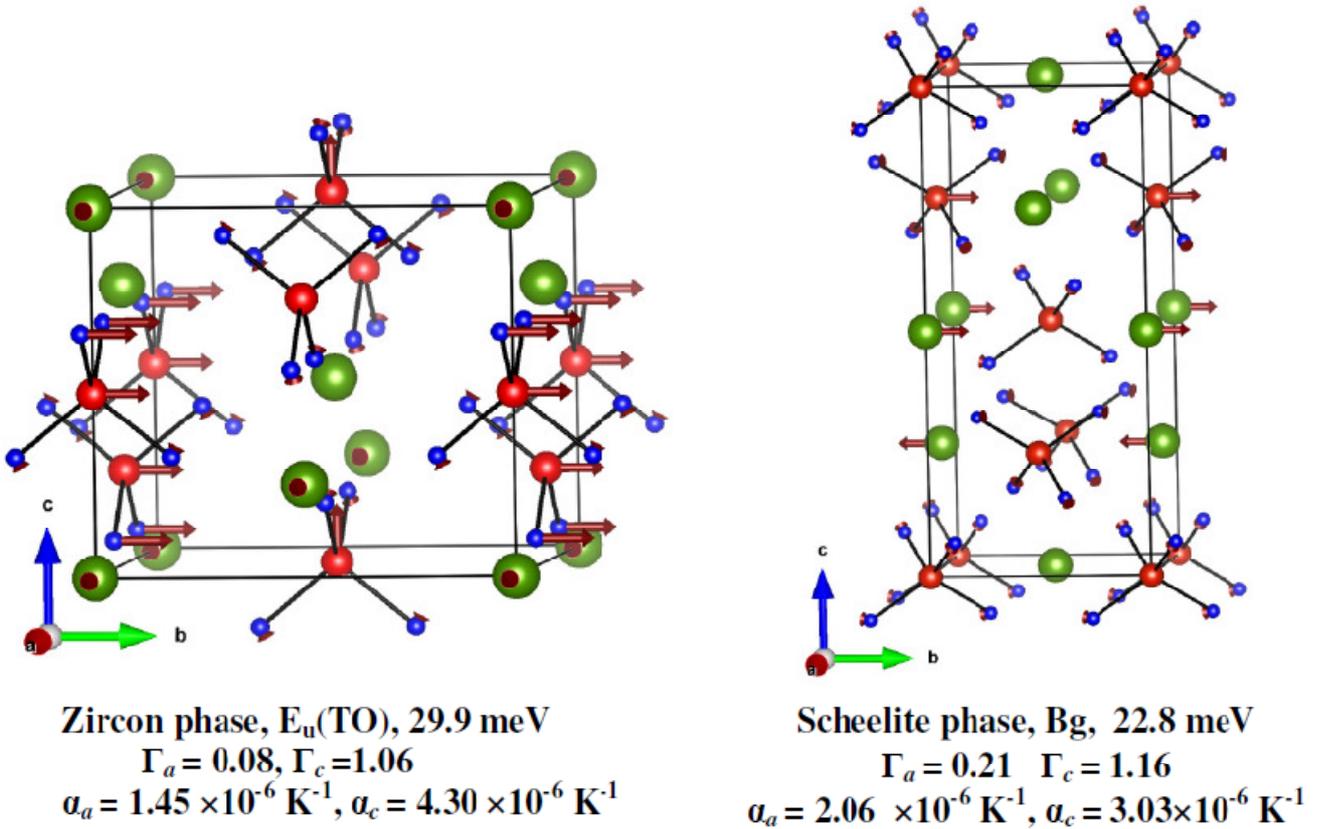

Zircon phase, $E_u$(TO), 29.9 meV
$\Gamma_a = 0.08$, $\Gamma_c = 1.06$
$\alpha_a = 1.45 \times 10^{-6}$ K$^{-1}$, $\alpha_c = 4.30 \times 10^{-6}$ K$^{-1}$

Scheelite phase, B$_g$, 22.8 meV
$\Gamma_a = 0.21$  $\Gamma_c = 1.16$
$\alpha_a = 2.06 \times 10^{-6}$ K$^{-1}$, $\alpha_c = 3.03 \times 10^{-6}$ K$^{-1}$